# The Implications of TeV-detected GRB Afterglows for Acceleration at Relativistic Shocks

Zhi-Qiu Huang[1], John G. Kirk[1], Gwenael Giacinti[1,2,3], and Brian Reville[1]
[1] Max-Planck-Institut für Kernphysik, Postfach 10 39 80, D-69029 Heidelberg, Germany
[2] Tsung-Dao Lee Institute, Shanghai Jiao Tong University, Shanghai 200240, People's Republic of China
[3] School of Physics and Astronomy, Shanghai Jiao Tong University, Shanghai 200240, People's Republic of China

## Abstract

Motivated by the detection of very-high-energy (VHE) gamma rays deep in the afterglow emission of a gamma-ray burst (GRB), we revisit predictions of the maximum energy to which electrons can be accelerated at a relativistic blast wave. Acceleration at the weakly magnetized forward shock of a blast wave can be limited by either the rapid damping of turbulence generated behind the shock, the effect of a large-scale ambient magnetic field, or radiation losses. Within the confines of a standard, single-zone, synchrotron self-Compton (SSC) model, we show that observations of GRB 190829A rule out a rapid damping of the downstream turbulence. Furthermore, simultaneous fits to the X-ray and TeV gamma-ray emission of this object are not possible unless the limit on acceleration imposed by the ambient magnetic field is comparable to or weaker than that imposed by radiation losses. This requires the dominant length scale of the turbulence behind the shock to be larger than that implied by particle-in-cell simulations. However, even then, Klein–Nishina effects prevent production of the hard VHE gamma-ray spectrum suggested by observations. Thus, TeV observations of GRB afterglows, though still very sparse, are already in tension with the SSC emission scenario.

*Unified Astronomy Thesaurus concepts:* High energy astrophysics (739); Gamma-ray bursts (629)

## 1. Introduction

The spectral evolution of the afterglows of gamma-ray bursts (GRBs) provides crucial information on their properties, as well as those of the surrounding medium (Mészáros 2002; Piran 2004; Kumar & Zhang 2015). To date, these data have predominantly been modeled under the assumption that the radiating particles are accelerated at or close behind the relativistic shock front that the GRB drives into its surroundings, although an alternative scenario in which electrons are injected via postburst activity in the central engine has its attractions (Kirk et al. 2021). When predicting the spectra and the characteristic frequencies of afterglow radiation, theoretical uncertainties have generally been parameterized by three essential quantities: the power-law index ($p$) of the accelerated particles and the ratio of the energy density in the accelerated electrons ($\epsilon_e$) and magnetic field ($\epsilon_B$) to that of the shocked fluid. The external shock acceleration model has long been the preferred scenario due not only to the relative simplicity of the model but also because the inferred injection spectra are broadly consistent with predictions from shock acceleration theory (e.g., Kirk et al. 2000).

Since the afterglow is observed mainly in the radio, optical, and X-ray bands, there has been relatively little interest in predicting the maximum energy to which particles can be accelerated. However, recent observations of afterglow radiation in very-high-energy (VHE) $\gamma$ rays by the H.E.S.S. and MAGIC collaborations (Abdalla et al. 2019, 2021; Acciari et al. 2019) have prompted us to reconsider this question.

In a previous paper, Kirk & Reville (2010, hereafter KR), we estimated the maximum energy to which particles can be accelerated and the frequency of the photons they radiate without considering in detail the dynamics of the shock. For ultrarelativistic shocks, we found that the maximum synchrotron photon energy falls well below the synchrotron burn-off limit (see also Derishev 2007; Lemoine 2013; Sironi et al. 2013; Asano et al. 2020). Here we compute the characteristic frequencies of the synchrotron and related synchrotron self-Compton (SSC) radiation from a GRB afterglow under the assumption of self-similar evolution of the shock front (Blandford & McKee 1976) and develop a simple single-zone ("thin shell") model to determine the spectrum and light curves. We revise expectations regarding different limitations on the maximum electron energy and critically assess their applicability in light of TeV $\gamma$-ray observations.

The outline of the paper is as follows. In the next section, we provide an overview of maximum-energy predictions for shock-accelerated electrons at weakly magnetized ultrarelativistic shocks. In Section 3, the predicted photon spectra of a self-similar blast wave are presented and compared to recent GRB detections by the H.E.S.S. experiment. The implications of our results for the shock acceleration model and the potential of future observations to sharpen these conclusions are discussed in Section 4.

## 2. The Maximum Energy of Accelerated Electrons

The Lorentz factor of the decelerating outer shock in the self-similar model of Blandford & McKee (1976) satisfies $\Gamma_{sh} \propto t^{-m/2}$, the two principal cases of interest being shock propagation into a uniform medium ($m=3$) or a wind profile ($m=1$). Here $t$ is time measured in the rest frame of the explosion's progenitor. To compare radiation signatures with observations, relevant parameters should be expressed in terms of the "observer time", $t_{obs} = (1+z)(t - \mathbf{n} \cdot \mathbf{r}/c)$, with $z$ the redshift of the source, $\mathbf{r}$ the position vector of the emitting plasma relative to the site of the explosion, and $\mathbf{n}$ the observer's direction. Since the radial velocity of the plasma is highly relativistic, those parts of the shock moving close to the direction of the observer make the dominant contribution, and the problem can be simplified by defining $t_{obs}$ as the time at which a photon emitted at the point on the shock front closest to Earth reaches the observer, $t_{\rm obs} \approx (1+z)t/[2(m+1)\Gamma_{\rm sh}^2]$ (this simplification can, of course, be relaxed; e.g., Panaitescu & Mészáros 1998; Chevalier & Li 2000).

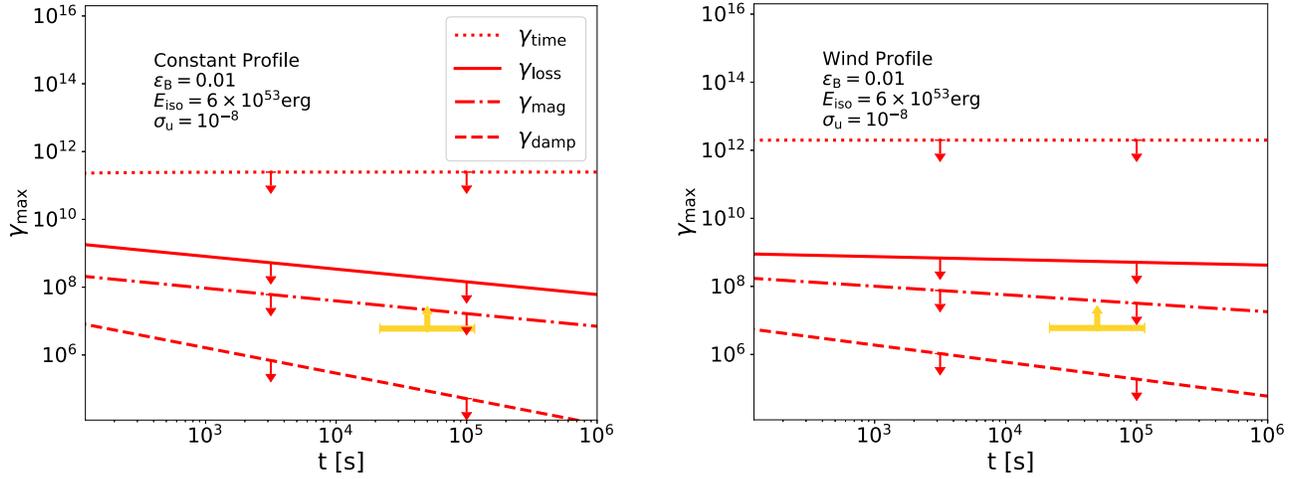

**Figure 1.** Maximum electron energy as a function of observer time for different external profiles. Left panel: constant-density profile. Right panel: wind density profile. From bottom to top, we identify four different limits: damping limited (dashed), magnetization limit (dashed–dotted), radiation loss limited (solid), and time limited (dotted). The maximum detected photon energy serves as a lower limit on the maximum electron energy in the source and is shown in yellow for the late-time VHE emission of GRB 190829A. The damping limit is shown for $L_{\rm damp} \approx \sigma_u^{-1/2} c/\omega_p$ (see text), and the loss and magnetization limits for $\ell_w = 10$.

We consider the adiabatic expansion of a blast wave of total energy $E$ into (a) a uniform medium of number density $n$ ($m=3$) and (b) a medium with number density $n = A/r^2$ ($m=1$). In terms of observer time and to lowest order in $\Gamma_{\rm sh}^{-2}$, the shock Lorentz factor decelerates as

$$\Gamma_{\rm sh} = \begin{cases} \left[\frac{17(1+z)^3 E}{4096\pi n m_p c^5}\right]^{1/8} t_{\rm obs}^{-3/8} & \text{case (a)} \\ \left[\frac{9(1+z)E}{32\pi A m_p c^3}\right]^{1/4} t_{\rm obs}^{-1/4} & \text{case (b),} \end{cases} \quad (1)$$

where $m_p$ is the proton mass, and we consider only a pure electron–proton external medium. We will assume the electron acceleration time is short compared to the dynamical time of the shock (verified in Figure 1), such that the maximum energy due to losses or other effects can be determined based on the instantaneous shock conditions.

Predictions for the maximum electron energy in a decelerating blast wave follow from the improved understanding of the microphysics of ultrarelativistic weakly magnetized ($\sigma_u \ll 1$) shocks provided by kinetic plasma simulations. Here $\sigma_u = B_1^2/4\pi n_1 m_p c^2$ is the upstream magnetization parameter, where $n_1$ and $B_1$ are, respectively, the proper number density and magnetic field strength immediately upstream of the shock.

The key parameter used by KR to characterize the downstream turbulence is its "strength", $a$, that is, the ratio of the characteristic size $\lambda$ of the intense magnetic field structures that mediate the shock transition (and persist downstream), to the length scale defined by the averaged turbulent magnetic field strength:

$$a = \lambda[eB/(m_e c^2)]. \quad (2)$$

Here $B$ is the rms magnetic field strength inside the structures, and $m_e$ ($-e$) is the mass (charge) of the electron. Kinetic simulations indicate that the typical size of the structures is several plasma inertial lengths. Therefore, we define $\lambda = \ell_w c/\omega_p$, where $\omega_p$ is the plasma frequency of the relativistic fluid, and follow Sironi et al. (2013) by setting $\ell_w = 10$ in our modeling, though values as large as 30 are also quoted. The effective mass of the particles in the immediate postshock relativistic proton–electron fluid is determined by the pressure $p_2 = (2/3)\Gamma_{\rm sh}^2 n_1 m_p c^2$ and the proper particle number density $n_2 = 2\sqrt{2}\,\Gamma_{\rm sh} n_1$, leading to $\omega_p = \sqrt{\pi n_2^2 e^2 c^2/p_2}$ (Amano & Kirk 2013), where we have used the jump conditions for a cold upstream medium. Simulations of weakly magnetized relativistic shocks show the magnetization parameter associated with the self-generated field $\epsilon_B = B^2/(24\pi p_2)$ peaks at the shock's position. Strong fluctuating fields extend upstream and downstream but plateau downstream at a level of $10^{-3}$–$10^{-2}$ (e.g., Keshet et al. 2009; Vanthieghem et al. 2020). The plateau value is important for the determination of the maximum electron energy.

In terms of the postshock magnetization parameter, one finds in the immediate downstream region of the self-similar expanding shock

$$a = \begin{cases} 2.6 \times 10^4 \ell_w \epsilon_B^{1/2} E_{54}^{1/8} n_1^{-1/8} t_{10}^{-3/8}(1+z)^{3/8} & \text{case (a)} \\ 5.7 \times 10^4 \ell_w \epsilon_B^{1/2} E_{54}^{1/4} A_{35}^{-1/4} t_{10}^{-1/4}(1+z)^{1/4} & \text{case (b),} \end{cases}$$
$$\quad (3)$$

where $t_{10} = t_{\rm obs}/10$ hr, and $E_{54} = E/10^{54}$ erg. In case (b), the upstream density is determined by the (assumed) constant mass-loss rate of the progenitor and its asymptotic wind velocity. We use $A = A_{35} \times 10^{35}$ cm$^{-1}$, corresponding to a mass-loss rate of $\dot{M} = 10^{-5} M_\odot$ yr$^{-1}$ and wind velocity $v_{\rm wind} = 3000$ km s$^{-1}$. From Equation (3), it is clear that $a \gg 1$, which implies that the electron emits synchrotron radiation (Landau & Lifshitz 1971) rather than "jitter" radiation, although low-frequency modifications due to repeated scatterings may nevertheless be present (Fleishman 2006; Reville & Kirk 2010).

The parameter $a$ also determines the transport regime for accelerated electrons; those with Lorentz factor $\bar{\gamma}_e < a$ (hereafter, a bar denotes quantities measured in the frame comoving with the plasma) are tied to magnetic field lines and undergo "helical" transport between scatterings. On the other hand, those with $\bar{\gamma}_e > a$ are deflected by only a small angle per fluctuation interaction and undergo "ballistic" transport between scatterings. In each case, the turbulence is expected to drive the particle distribution toward isotropy. However, in doing so, it causes the electrons to radiate. In relativistic shock acceleration theory, the isotropization rate is thought to be roughly equal to the rate of energy gain. In each transport regime, one can define a Lorentz factor $\bar{\gamma}_{max}$ such that isotropization/energization is slower than radiation losses if $\bar{\gamma}_e > \bar{\gamma}_{max}$. Generalizing the results found by KR to include cooling by inverse Compton scattering, one finds the maximum energy when limited by radiative losses

$$\bar{\gamma}_{loss} = \begin{cases} a_{crit} & a < a_{crit} \\ a_{crit}\sqrt{a_{crit}/a} & a > a_{crit} \end{cases}, \quad (4)$$

where

$$a_{crit} = \{3m_e \ell_w c^3/[2e^2\omega_p(1+x)]\}^{1/3}$$
$$= \begin{cases} 4.1 \times 10^6 \ell_w^{1/3} n_1^{-1/6}(1+x)^{-1/3} & \text{case (a)} \\ 8.9 \times 10^6 E_{54}^{1/6} \ell_w^{1/3} A_{35}^{-1/3} & \\ \times t_{10}^{1/6}(1+z)^{-1/6}(1+x)^{-1/3} & \text{case (b).} \end{cases} \quad (5)$$

Here $x = \dot{\gamma}_{IC}/\dot{\gamma}_{syn}$ is the ratio of the power emitted by a single electron as inverse Compton emission to that emitted as synchrotron radiation. If Compton scattering takes place in the Thomson regime, $x$ is independent of the electron energy. Furthermore, neglecting the contribution of external photons, $x \approx \bar{U}_{syn}/\bar{U}_B$, the ratio of the energy density of synchrotron photons to that of the magnetic field, measured in the comoving frame (see Section 3.3). Comparing Equations (3) and (5), we expect $a \ll a_{crit}$, which implies that transport of the highest-energy electrons proceeds in the ballistic regime via small-angle scattering.

In this regime, the isotropization rate is (KR)

$$\nu_{iso} \approx \langle \Delta\theta^2 \rangle \nu_{sc} = \left(\frac{a}{\bar{\gamma}}\right)^2 \frac{c}{\lambda}. \quad (6)$$

Although isotropization is dominated by frequent interactions with small-scale fields, the rate at which this occurs decreases rapidly with increasing Lorentz factor. At sufficiently large energies, the effect on a particle trajectory becomes negligible compared to that of the magnetic field averaged over scales much larger than the scattering mean free path, $c/\nu_{iso}$. This large-scale field downstream of the shock can be determined from the ambient magnetic field via the shock jump conditions, $\langle B_{2,\parallel} \rangle = \langle B_{1,\parallel} \rangle$ and $\langle B_{2,\perp} \rangle = 2\sqrt{2}\Gamma_{sh}\langle B_{1,\perp} \rangle$, the average being taken over spatial scales $\gg \lambda$. Defining the mean downstream electron gyrofrequency $\omega_g = e\langle B_2 \rangle/\bar{\gamma} m_e c$, electrons are said to be magnetized when $\omega_g > \nu_{iso}$. Unless the upstream field is aligned with the shock normal to within $1/\Gamma_{sh}$ (a situation we do not consider here), the downstream mean field is effectively perpendicular to the bulk flow direction. Achterberg et al. (2001) found that for acceleration to proceed in this case, the isotropization rate in the downstream frame should exceed the gyrofrequency in the average field if particles are to overtake the receding shock (see also Lemoine & Pelletier 2010). For a shock with an external magnetization parameter $\sigma_u = B_1^2/4\pi n_1 m_p c^2$, and as before, turbulent magnetic fluctuations concentrated at $\lambda = \ell_w c/\omega_p$, it follows that particles are magnetized downstream when their Lorentz factor exceeds

$$\bar{\gamma}_{mag} = \ell_w \frac{m_p}{m_e} \epsilon_B \sigma_u^{-1/2}. \quad (7)$$

The maximum electron Lorentz factor is then determined by the more stringent of the two limits: $\bar{\gamma}_{max} = \text{Min}(\bar{\gamma}_{loss}, \bar{\gamma}_{mag})$. We note that additional large-scale magnetic field fluctuations related to upstream wave excitation (Reville & Bell 2014) or ambient turbulence may provide a channel to circumvent the magnetized energy limit, though this may introduce other spectral features (e.g., Niemiec et al. 2006).

In addition to the loss and magnetized limits, a third constraint on the maximum electron Lorentz factor arises if the magnetic field fluctuations generated at the shock damp away rapidly (see also Lemoine 2013). Defining $L_{damp}$ as the characteristic length scale of this decay, it follows that the return probability for particles with a mean free path $c/\nu_{sc} > L_{damp}$ is significantly reduced. Simulations by Sironi et al. (2013), for example, display a symmetric rise and decay profile of $\epsilon_B$ about the shock, with a characteristic scale[4] $L_{damp} \approx \sigma_u^{-1/2} c/\omega_p$. The maximum energy is therefore limited by this effect to

$$\bar{\gamma}_{damp} \approx (\ell_w \epsilon_B m_p/m_e)^{1/2} \Gamma_{sh} \sigma_u^{-1/4}, \quad (8)$$

which we refer to as the damping limit.

Figure 1 shows the three limits on the maximum electron Lorentz factor given by Equations (4), (7), and (8), taking instantaneous values of the shock parameters in the two scenarios considered. Standard GRB afterglow parameters are adopted here and in subsequent figures. The values are boosted to the upstream observer's frame for comparison to observations (neglecting redshift). The detection of late-time TeV emission from two GRBs places a firm lower limit on the maximum particle energy in the observer's frame, indicated in yellow in the figure. As a consistency check, we also indicate the maximum energy that might be expected from time-limited acceleration using the prescription of Achterberg et al. (2001),

$$\bar{\gamma}_{time} = \bar{\gamma}(t_0) + \int_{t_0}^{t_{obs}} dt \frac{\bar{\gamma}(t)}{t_{cyc}(\bar{\gamma}, t)}, \quad (9)$$

where $t_{cyc}$ is the cycle time assuming ballistic scattering only in the downstream and the combined action of scattering and regular deflection in the upstream. We choose as an initial condition $\bar{\gamma}(t_0) = \Gamma_{sh}(t_0)$, and since $\Gamma_{sh}$ diverges as $t \to 0$, we select $t_0$ such that $\Gamma_{sh}(t_0) = 300$. This equation is integrated numerically from $t_0$ to the observed time to give the time-limited acceleration upper limit in Figure 1. The fact that this limit substantially exceeds the others justifies our use of instantaneous flow parameters when evaluating the latter.

## 3. The Synchrotron Self-Compton Spectrum

### 3.1. Characteristic Frequencies

In the context of the rough estimates we seek here, synchrotron radiation can be treated as narrowband; i.e., electrons with a Lorentz factor $\bar{\gamma}_e$ radiate only photons close to a characteristic frequency given by $\bar{\nu} = h_1 B \bar{\gamma}_e^2$, where $h_1 = 1.3 \times 10^6$ Hz G$^{-1}$. This frequency is independent of pitch angle, and the photons can be considered to be radiated isotropically in the comoving frame of the plasma. At the observer, these photons appear blueshifted by the Doppler factor $\mathcal{D} = \gamma_2(1 + \beta\bar{\mu})$, where $\gamma_2 = (1 - \beta^2)^{-1/2} = \Gamma_{\rm sh}/\sqrt{2}$ is the bulk Lorentz factor of the ultrarelativistic expanding plasma immediately downstream of the shock front, and $\bar{\mu}$ is the cosine of the angle between the photon momentum and the radius vector $\boldsymbol{r}$ as measured in the comoving frame. Also taking the cosmological redshift $z$ into account, the highest-frequency photons that can be produced by shock acceleration emerge from the closest point to Earth on the shock (i.e., $\bar{\mu} = 1$, $\mathcal{D} \approx 2\gamma_2$),

$$\nu_{\max}^{\rm loss} = \mathcal{D} h_1 B a_{\rm crit}^2/(1+z)$$
$$= \begin{cases} 0.6\ell_{\rm w}^{2/3}\epsilon_{B,-2}^{1/2} n_1^{-1/12} E_{54}^{1/4} t_{10}^{-3/4} \\ \times (1+z)^{-1/4}(1+x)^{-2/3} \text{ MeV} \quad \text{case (a)} \\ 1.2\ell_{\rm w}^{2/3}\epsilon_{B,-2}^{1/2} E_{54}^{1/3} A_{35}^{-1/6} t_{10}^{-2/3} \\ \times (1+z)^{-1/3}(1+x)^{-2/3} \text{ MeV} \quad \text{case (b),} \end{cases} \quad (10)$$

where $\epsilon_B = 10^{-2}\epsilon_{B,-2}$. Here we have assumed ballistic transport. If helical transport applies, i.e., $a > a_{\rm crit}$, the maximum synchrotron photon energy is $h\nu_{\max}^{\rm loss}/m_e c^2 = \mathcal{D}(1+x)^{-1}\alpha_f^{-1}$, where $\alpha_f$ is the fine-structure constant. If, on the other hand, magnetization of the electrons limits their acceleration, Equation (7), then the synchrotron limit is

$$\nu_{\max}^{\rm mag} = \mathcal{D}h_1 B \ell_w^2 \left(\frac{m_p}{m_e}\right)^2 \epsilon_B^2 \sigma_u^{-1}/(1+z)$$
$$= \begin{cases} 1.1\,\ell_{\rm w}^2 \sigma_{-8}^{-1} \epsilon_{B,-2}^{5/2} n_1^{1/4}(1+z)^{-1/4} E_{54}^{1/4} t_{10}^{-3/4} \text{keV} & \text{case (a)} \\ 0.5\,\ell_{\rm w}^2 \sigma_{-8}^{-1} \epsilon_{B,-2}^{5/2} A_{35}^{1/2} t_{10}^{-1} \text{ keV} & \text{case (b)} \end{cases}, \quad (11)$$

where $\sigma_u = 10^{-8} \times \sigma_{-8}$. Note here the strong dependence on $\ell_w$ and $\epsilon_B$, in contrast to the cooling-limited maximum frequency. For example, if $\epsilon_B \ll 0.01$, the maximum synchrotron photons may not reach the X-ray band at late times.

---
[4] Approximately equal to the upstream gyroradius of shock-reflected particles as measured by a downstream observer.

We emphasize that in both cases, the synchrotron cutoff lies below the frequently quoted maximum value (synchrotron burn-off/Bohm limit) of $2\gamma_2 m_e c^2/\alpha_f = 2\gamma_2 \times 68$ MeV, where $\alpha_f$ is the fine-structure constant. Employing a multizone emission model to overcome the synchrotron burn-off limit (e.g., Kumar et al. 2012; Khangulyan et al. 2021) must also contend with the maximum-energy predictions for weakly magnetized shocks.

The lower bound on the Lorentz factor of accelerated electrons is conventionally fixed by the efficiency with which energy is extracted from the particles flowing into the shock via the equation

$$\bar{\gamma}_{\min}\left[\frac{1-(\bar{\gamma}_{\min}/\bar{\gamma}_{\max})^{p-2}}{1-(\bar{\gamma}_{\min}/\bar{\gamma}_{\max})^{p-1}}\right] = \left(\frac{p-2}{p-1}\right)\frac{m_p}{m_e}\epsilon_e \gamma_2, \quad (12)$$

where $p$ is the index of the injection spectrum assumed to be a power law with abrupt low- and high-energy cutoffs (see Derishev & Piran 2021, for alternative approaches). It is commonly assumed that $\bar{\gamma}_{\max} \gg \bar{\gamma}_{\min}$, in which case the term in square brackets can be set to unity. However, special care should be taken if $p$ is close to 2, as suggested, for example, by Ajello et al. (2019). Although the corresponding correction to $\bar{\gamma}_{\min}$ is typically small, it has a significant impact on the resulting SSC emission (see Petropoulou et al. 2011).

The observation frequency corresponding to $\bar{\gamma}_{\min}$ (written in hertz, as it falls in the infrared band) is

$$\nu_{\min} = \mathcal{D} h_1 B \bar{\gamma}_{\min}^2/(1+z)$$
$$\approx \epsilon_{B,-2}^{1/2} \epsilon_{e,-1}^2 E_{54}^{1/2} t_{10}^{-3/2}$$
$$\times \sqrt{1+z}\left(\frac{p-2}{p-1}\right)^2 \begin{cases} 22.7 \text{ THz} & \text{case (a)} \\ 45.7 \text{ THz} & \text{case (b),} \end{cases} \quad (13)$$

where we have again selected $\bar{\mu} = 1$ and ignored the correction in Equation (12) due to finite maximum energy.

Once injected, an accelerated electron loses energy primarily via inverse Compton and synchrotron radiation in the downstream plasma, and its Lorentz factor $\bar{\gamma}_e$ changes at the rate $\dot{\bar{\gamma}}_{\rm rad}(\bar{\gamma}_e) = -(1+x)h_0 \bar{\gamma}_e^2 B^2$, where $h_0 = 1.29 \times 10^{-9}$ s$^{-1}$ G$^{-2}$. Provided $\bar{\gamma}_e/|\dot{\bar{\gamma}}_{\rm rad}(\bar{\gamma}_e)|$ is shorter than the dynamical evolution time of the shell, $\bar{t}_{\rm dyn} \approx t/2\gamma_2$, i.e., provided $\bar{\gamma}_e > \bar{\gamma}_c = 2\gamma_2/[(1+x)h_0 B^2 t]$, the particles lose essentially all of their energy to radiation, the synchrotron component of which is observed at frequencies greater than the "cooling frequency":

$$\nu_c = \mathcal{D} h_1 B \bar{\gamma}_c^2/(1+z)$$
$$\approx \begin{cases} 48\epsilon_{B,-2}^{-3/2} n_1^{-1} E_{54}^{-1/2} t_{10}^{-1/2} \frac{(1+z)^{-\frac{1}{2}}}{(1+x)^2} \text{ THz} & \text{case (a)} \\ 9.7\epsilon_{B,-2}^{-3/2} E_{54}^{1/2} A_{35}^{-2} t_{10}^{1/2} \frac{(1+z)^{-\frac{3}{2}}}{(1+x)^2} \text{ PHz} & \text{case (b).} \end{cases} \quad (14)$$

If the loss rate is such that $\nu_c > \nu_{\min}$ or, equivalently, $\bar{\gamma}_c > \bar{\gamma}_{\min}$, the emission is said to occur in the slow-cooling regime, since the bulk of the electrons lose only a fraction of their energy to radiative cooling.

### 3.2. Synchrotron Spectrum

Apart from very early times, immediately after the prompt phase, it is clear from Equations (13) and (14) that, for typical GRB parameters, $\nu_c > \nu_{\min}$; i.e., synchrotron emission occurs in the slow-cooling regime. It is common practice in this case to neglect the emission from electrons that were injected more than a dynamical time earlier, although this procedure is not always justified (see, for example, Granot et al. 1999). The emitting electrons are thus assumed to occupy a thin homogeneous shell immediately behind the shock front and produce a spectral power that we approximate as $F(\nu) \propto \nu^{1/3}$ for $\nu < \nu_{\min}$, $F(\nu) \propto \nu^{-(p-1)/2}$ for $\nu_{\min} < \nu < \nu_c$, and $F(\nu) \propto \nu^{-p/2}\exp(-\nu/\nu_{\max})$ for $\nu > \nu_c$.

Now consider an element of this shell of area $dA$, which, in the comoving frame, emits synchrotron radiation with spectral power

$$\frac{d\bar{F}(\bar{\nu})}{dA} = m_e c^2 |\dot{\bar{\gamma}}_{\rm syn}| \frac{dn}{d\bar{\gamma}_e} \frac{d\bar{\gamma}_e}{d\bar{\nu}}, \quad (15)$$

where $\dot{\bar{\gamma}}_{\rm syn} = \dot{\bar{\gamma}}_{\rm rad}/(1+x)$ is the cooling rate due to synchrotron losses only. For $\bar{\gamma}_e > \bar{\gamma}_c$, a steady-state electron distribution, $dn/d\bar{\gamma}_e$ (per unit area of the shock), is established such that the rate $\dot{n}_>(\bar{\gamma}_e)$ at which particles are injected with Lorentz factors greater than $\bar{\gamma}_e$ equals the rate at which they cool to lower Lorentz factors, $|\dot{\bar{\gamma}}_{\rm rad}(\bar{\gamma}_e)| dn/d\bar{\gamma}_e$. The spectral power in this range is, therefore,

$$\frac{d\bar{F}(\bar{\nu} > \bar{\nu}_c)}{dA} = \frac{m_e c^2}{2h_1 B \bar{\gamma}_e} \frac{\dot{n}_>(\bar{\gamma}_e)}{1+x}, \quad (16)$$

where it is understood that electrons at a given Lorentz factor emit at $\bar{\gamma}_e = \sqrt{h_1 B/\bar{\nu}}$.

On the other hand, for $\bar{\gamma}_e < \bar{\gamma}_c$, a steady state is achieved because particles are assumed to cease radiating a certain time $\bar{t}_{esc}$ after injection without having appreciably changed their Lorentz factor, i.e., $|d\dot{n}_>/d\bar{\gamma}_e| = \bar{t}_{esc}^{-1} dn/d\bar{\gamma}_e$. In this case, the spectral power either vanishes (for fast cooling) or is given by

$$\frac{d\bar{F}(\bar{\nu} < \bar{\nu}_c)}{dA} = \frac{m_e c^2}{2h_1 B \bar{\gamma}_e} |\dot{\gamma}_{syn}| \bar{t}_{esc} \left| \frac{d\dot{n}_>}{d\bar{\gamma}_e} \right|. \quad (17)$$

We exploit the freedom inherent in the single-zone model by choosing $\bar{t}_{esc} = \bar{t}_{dyn}/(p-1)$, which ensures that, in the slow-cooling case, the model flux is continuous at $\bar{\nu} = \bar{\nu}_c$. The peak spectral power (i.e., the maximum of $d\bar{F}(\bar{\nu})/dA$) for this shell element is emitted at $\bar{\nu}_{peak} = \min[\bar{\nu}_c, \bar{\nu}_{min}]$, at which point $\dot{n}_> = 4\gamma_2 n_1 c/3$ and, for slow cooling, $(p-1)^{-1}\bar{\gamma}_e|d\dot{n}_>/d\bar{\gamma}_e| = \dot{n}_>$. In each case,

$$\left.\frac{dn}{d\bar{\gamma}_e}\right|_{peak} = \frac{2n_1 R}{3\bar{\gamma}_{peak}}. \quad (18)$$

To relate the above spectral power to observations, we first note that it is emitted isotropically in the comoving frame. Denoting the direction to the observer by the unit vector $\bar{\Omega}$ in this frame, the spectral power per unit solid angle is $d\bar{F}/d\bar{\Omega} = \bar{F}/4\pi$. Then, in the observer's frame, the received spectral power per unit solid angle for a given area on the shock surface is $dF/d\Omega = \mathcal{D}^3 d\bar{F}/d\bar{\Omega}$ (Rybicki & Lightman 1986, Equation (4.97)). We simplify the integration over the shock surface by replacing $\nu_{peak}$, $B$, $n_1$, and $R$ with their values at $\bar{\mu} = 1$ and noting that $\int dA \mathcal{D}^3 = 4\pi R^2 \gamma_2$. With these simplifications, one finds the peak flux density at the observer

$$\mathcal{F}^{SYN}(\nu_{peak}) = \frac{1}{D_L^2} \int \frac{dF(\nu_{peak})}{d\Omega} d\Omega_{shell} \quad (19)$$

$$= \frac{1}{D_L^2} \frac{h_0 m_e c^2}{3h_1} B n_1 R^3 \gamma_2, \quad (20)$$

where $D_L$ is the luminosity distance. For $x = 0$, this expression agrees to an order of magnitude with previous results (e.g., Sari et al. 1998).

### 3.3. Inverse Compton Emission

The effect of cutoffs in the maximum electron energy at modest values was previously demonstrated by Petropoulou et al. (2011), who explored the impact on both the light curves and spectra. Here we revisit this effect for the competing maximum-energy models discussed in Section 2.

We assume that the cooling break $\nu_c$ is determined by the combination of synchrotron and SSC losses in the Thomson regime. For consistency, this requires $\bar{\gamma} \ll \bar{\gamma}^* \equiv m_e c^2/h\bar{\nu}^*$, where $\bar{\nu}^* = \max[\bar{\nu}_{min}, \bar{\nu}_c]$. For typical GRB afterglow parameters, we find $\bar{\gamma}^* > \bar{\gamma}_c$. Rather extreme conditions are required for Klein–Nishina (KN) effects to significantly modify the electron spectrum (see Abdalla et al. 2021, Supp. Mat.).

However, as we show below, these effects have a strong impact on the high-energy photon spectrum.

To compute the SSC emission, we first need to model the specific energy density of synchrotron photons $\bar{U}(\bar{\nu})$ within the source. The specific intensity of radiation on the source surface is directly detectable, $\bar{I}_{\bar{\nu}} = d\bar{F}(\bar{\nu})/dA/d\bar{\Omega}$, but modeling this quantity in the interior of the source requires additional assumptions. Here we follow previous treatments and assume that $\bar{I}_{\bar{\nu}}$ within the source is homogeneous and isotropic, so that $\bar{U}(\bar{\nu}) = (\alpha/c) d\bar{F}(\bar{\nu})/dA$, and set the geometrical factor $\alpha$ to unity.[5] Inserting the synchrotron photon power given in Equations (16) and (17) then leads to an implicit expression for $x = \int \bar{U}(\bar{\nu}) d\bar{\nu}/\bar{U}_B$, which, in the limit $\bar{\gamma}_{max} \gg \bar{\gamma}_{min}$, reduces to an expression similar to Sari & Esin (2001, Equation (3.1)). However, for a general $\bar{\gamma}_{max}$, we solve the full expression for $x$ numerically. For simplicity, we again restrict ourselves to the Thomson regime, although the method can, in principle, be extended to include KN effects (in which case, $x$ is energy dependent). The resulting SSC flux density is (cf. Rybicki & Lightman 1986, Equation (7.28a))

$$\mathcal{F}^{SSC}(\nu_{IC}) = 3\sigma_T \int_{\bar{\gamma}_1}^{\bar{\gamma}_{max}} d\bar{\gamma}_e \frac{dn}{d\bar{\gamma}_e} \left(1 - \frac{h\nu_{IC}}{\mathcal{D}\bar{\gamma}mc^2}\right)$$
$$\times \int_{1/4\bar{\gamma}^2}^{1} dz\, g(z) \mathcal{F}^{SYN}\left(\frac{\nu_{IC}}{4\bar{\gamma}^2 z\left(1 - \frac{h\nu_{IC}}{\mathcal{D}\bar{\gamma}mc^2}\right)}\right),$$

where $g(z) = 1 + z + 2z\ln(z) - 2z^2 + \frac{1-z}{2}\left[\frac{(h\nu_{IC}/\mathcal{D}\bar{\gamma}mc^2)^2}{(1-h\nu_{IC}/\mathcal{D}\bar{\gamma}mc^2)}\right]$ (Jones 1968). The full KN expression should be used here, since an accurate picture at the highest photon energies is required. Adopting a step function cutoff in the synchrotron flux density above $\bar{\nu}_{max}$, the integral over $z$ can be performed analytically, leaving a single integration over the electron distribution, which is done numerically. The electron distribution follows from Equation (18) and the corresponding slopes of the synchrotron spectrum. In deriving the above, we have assumed isotropy of the electrons and photons in the fluid frame and exploited the fact that $\mathcal{F}^{SSC}$ and $\mathcal{F}^{SYN}$ transform in the same way (Sari & Esin 2001).

In Figure 2, we compare representative fits for the wind and constant-density scenarios to the afterglow emission of the nearest VHE-detected GRB (GRB 190829A), since it is the object least affected by uncertainties connected with intergalactic absorption that, nevertheless, has significant late-time detection (Abdalla et al. 2021). While several recent theoretical works on this have tried to match the observations in a standard SSC model, the maximum energy was either assumed to be at the Bohm limit (Lu-Lu et al. 2021) or left as a free parameter (Salafia et al. 2021). Here we attempt to quantify the extent to which observations are consistent with current maximum-energy predictions discussed in the previous sections.

For $\ell_w = 10$, the maximum electron energy is fixed by the magnetized limit (see Figure 1) and is too low to match either the X-ray or $\gamma$-ray data (see Figure 3). In addition, $\bar{\gamma}_{damp} < \bar{\gamma}_{mag}$ for all $\ell_w$. We conclude that damping of magnetic field fluctuations is slow and plays no role. Allowing $\ell_w$ to increase permits one to fit the X-ray data. However, irrespective of the maximum energy, KN suppression causes the SSC $\gamma$-ray spectra to be unavoidably softer than the data suggest (Abdalla et al. 2021), with $\nu F_\nu \sim \nu^{-0.23}$ at early times, steepening to $\sim \nu^{-0.33}$ at later times. The dependence of the synchrotron and SSC spectrum on $\ell_w$, and hence the maximum electron energy, is illustrated in Figure 3, where the spectrum is calculated for GRB 190829A at 5.5 hr after the prompt emission using a constant external density profile. We note that for $\ell_w = 10$, $\bar{\gamma}_{\rm mag} \ll \bar{\gamma}_{\rm loss}$; for $\ell_w = 100$, $\bar{\gamma}_{\rm mag} \approx \bar{\gamma}_{\rm loss}$; and for $\ell_w = 1000$, $\bar{\gamma}_{\rm mag} \gg \bar{\gamma}_{\rm loss}$. In the extreme $\ell_w = 10^4$ case, $a \approx a_{\rm crit}$ and is very close the synchrotron burn-off limit. While the synchrotron cutoff is pushed to higher frequencies with increasing $\ell_w$ (cf. Equations (10) and (11)), the SSC spectrum above 100 GeV unavoidably softens due to Klein–Nishina suppression. Detection of a turnover, or its absence, in hard X-rays (see, for example, Kouveliotou et al. 2013) in future TeV-detected afterglows would clearly constrain the models further.

---

[5] The SSC expression that follows reduces in the Thomson limit to that of Sari & Esin (2001) if we take $\alpha = (p-1) \approx 1$, due to our definition of $\bar{t}_{esc}$. Note that the $R$ in their Equation (A2) corresponds to the postshocked radius in which the emitting electrons reside, which, following Blandford & McKee (1976), is $\approx R/\Gamma_{sh}^2$ for an upstream observer in our notation.

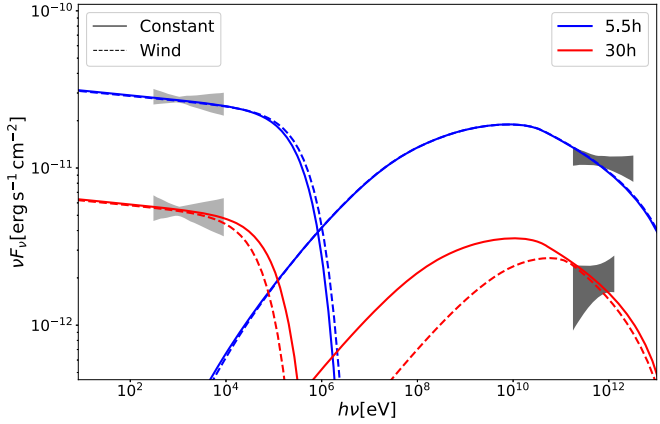

**Figure 2.** Model fit to GRB 190829A using the method described in the text (which rejects the potential damping limit). The luminosity distance is $D = 1 \times 10^{27}$ cm. For the wind density profile, $A_{35} = 2$, while for the constant profile, $n_1 = 1\,\text{cm}^{-3}$. In both cases, the explosion energy is $E = 5 \times 10^{52}$ erg, electron injection index $p = 2.06$, $\sigma_u = 10^{-9}$, and $\ell_w = 100$. The values of ($\epsilon_{e,-2}$, $\epsilon_{B,-3}$, $x_{IC}$, $\bar{\gamma}_{\min}$) are (5.7, 1.3, 7.4, 93) and (4.3, 1.3, 4, 35) at $t = 5.5$ and 30 hr, respectively, in the constant-density case and (5.8, 1.2, 8.6, 111) and (4.7, 1.0, 5.3, 58) in the wind profile scenario. The X-ray data are automatically processed by the UK Swift Science Data Centre (Evans et al. 2007, 2009), and the gamma-ray butterflies are from Abdalla et al. (2021).

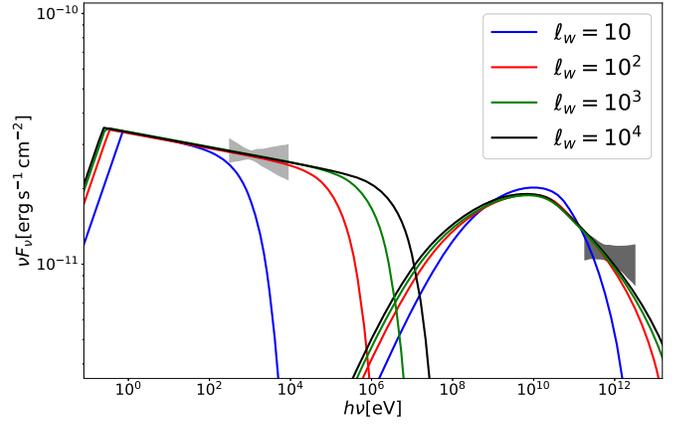

**Figure 3.** Comparison of predictions with varying $\ell_w$ for GRB 1908929A at $t = 5.5$ hr for the constant-density case. Environmental parameters are the same as used in Figure 2. Acceleration parameters are selected to match the low-energy synchrotron flux, with the SSC spectrum intersecting the H.E.S.S. butterfly at the upper leftmost point. The values of ($\epsilon_{e,-2}$, $\epsilon_{B,-3}$, $x_{IC}$, $\bar{\gamma}_{\min}$) are the same for $\ell_w = 100$ and (4, 1, 6, 78), (6.3, 1.4, 7.8, 98), and (7, 1.4, 8.4, 103) for $\ell_w = 10$, $10^3$, and $10^4$ respectively. See text for further details.

For fixed $\epsilon_e$ and $\epsilon_B$, the TeV flux in our model decays with observed time as $\propto t_{\rm obs}^{-s}$, with $s \approx 1.2$ in the uniform medium case and $s \approx 1.3$ in the wind case, both being slightly more rapid than the decay rate by the H.E.S.S. data. By admitting a temporal (or $\gamma_2$) dependence of $\epsilon_B$ and/or $\epsilon_e$, it is not difficult to produce a light curve that matches both the X-ray and TeV data. However, such a dependence is, to our knowledge, not seen in simulations.

## 4. Discussion

The detection of several GRB afterglows in the TeV $\gamma$-ray domain provides a critical test not only of the standard external shock model of GRB afterglows, but of the particle acceleration theory at weakly magnetized ultrarelativistic shocks in general. Here we have focused on the constraints such measurements may place on the maximum electron energy within the external shock acceleration framework for GRB afterglows. The physical conditions implied by numerical simulations of weakly magnetized relativistic shocks generally indicate that acceleration will proceed with an efficiency significantly below the frequently adopted Bohm rate. The maximum synchrotron photon energy is thus predicted to fall short of the theoretical maximum synchrotron burn-off limit at $h\nu_{\max} \approx 100\,\Gamma_{\rm sh}$ MeV.

Following the approach of Kirk & Reville (2010), we consider different physical scenarios, motivated by the results of current kinetic simulations. Within the limitations of our assumptions, we conclude the following.

1. Scenarios in which the magnetic field damps rapidly downstream of the shock are clearly ruled out in the shock acceleration picture (see Figure 1). On the other hand, a plateau (or, equivalently, a very slow decay on a scale $\gg \sigma_u^{-1/2} c/\omega_p$) in the downstream field intensity (e.g., Vanthieghem et al. 2020) can be accommodated.

2. Adopting a characteristic scale of $\lambda = 10 c/\omega_p$ for the scattering centers, we find that quenching by the large-scale magnetic field, the so-called magnetization limit, is operative, and this scenario is also ruled out by the data. Larger-scale structures are, therefore, required but not seen to develop in the currently available simulations. However, this may be a limitation of the spatial extent/short duration of these studies. Future simulations with a larger dynamical range, along with a deeper understanding of upstream wave excitation and the nature of preexisting ambient turbulence, will assist in clarifying this question (see Lemoine & Pelletier 2010; Reville & Bell 2014).

3. If we allow for $\ell_w > 10$, the magnetization limit increases, and quenching by radiation losses ultimately sets the upper limit. However, this scenario is also challenged by the data. The measurements by Abdalla et al. (2021) show that the VHE $\gamma$-ray spectrum is hard. However, Klein–Nishina suppression softens the spectrum in the VHE $\gamma$-ray band and presents a significant obstacle to simultaneously matching the X-ray and $\gamma$-ray data. Current observations thus appear to also be in conflict with the loss-limited scenario.

Taken together, these conclusions present a serious challenge to the external shock acceleration model for GRB afterglows.

However, they are based on data from a single GRB and also weakened by the numerous simplifying assumptions conventionally adopted in the single-zone external shock model. It is perhaps not surprising that the discovery of late-time VHE γ-ray emission presents a challenge to the simplest picture. Future detection with high-quality spectral and temporal data will be essential to resolve this issue.

Finally, we speculate on some requirements for alternative models. Given the many successes of the external shock framework in meeting GRB afterglow data, it is attractive to explore alternative scenarios that do not deviate drastically from the essential aspects of the model, in particular, the self-similar nature of the blast wave. One such possibility is to consider an external Compton scenario (Zhang et al. 2020, 2021). To avoid the same spectral softening that challenges the SSC model, the external target photons should be scattered in the Thomson regime. Motivating the required photon fields to reproduce the spectral and temporal features requires careful consideration. Looking beyond the standard external shock scenario, one could consider a model in which electrons are instead accelerated in the jet either via shear (Rieger & Duffy 2005) or inductive (Kirk et al. 2021) acceleration but nevertheless radiate in the postshock amplified fields. This two-zone picture is appealing, since the decoupling of the acceleration and emission processes avoids the synchrotron burn-off limit for maximum-energy synchrotron photons (e.g., Kirk & Giacinti 2017). Future γ-ray detection will be critical in terms of both motivating and distinguishing such alternative scenarios.

The authors thank Andrew Taylor for valuable discussions.

## ORCID iDs

Zhi-Qiu Huang https://orcid.org/0000-0002-9239-323X
John G. Kirk https://orcid.org/0000-0002-9859-0496
Gwenael Giacinti https://orcid.org/0000-0001-9745-5738
Brian Reville https://orcid.org/0000-0002-3778-1432